\documentclass[twocolumn,aps,prc,showpacs,superscriptaddress,floatfix]{revtex4-2}
\usepackage{rotating}
\usepackage{epsfig} 
\usepackage{hyperref}
\usepackage{color}
\usepackage{url}
\usepackage{multirow}

\usepackage{amsmath}
\usepackage{lineno}
\usepackage{graphicx}
\usepackage[caption=false]{subfig}

\graphicspath{ {./plot/}, {./image/} }
\DeclareGraphicsExtensions{ .pdf, .png}

\makeatletter
\renewcommand{\p@subsection}{}
\renewcommand{\p@subsubsection}{}
\makeatother

\begin{document}


\newcommand{\der}{\text{d}}
\newcommand{\rp}{\Psi_{\textsc{rp}}}

\newcommand{\qcd}{{\textsc{qcd}}}
\newcommand{\pos}{{\textsc{os}}}
\newcommand{\pss}{{\textsc{ss}}}
\newcommand{\cme}{{\textsc{cme}}}
\newcommand{\ampt}{{\textsc{ampt}}}
\newcommand{\hijing}{{\textsc{hijing}}}
\newcommand{\poi}{{\textsc{poi}}}
\newcommand{\snn}{\sqrt{s_{\textsc{nn}}}}
\newcommand{\srp}{{\textsc{rp}}}
\newcommand{\ssp}{{\textsc{sp}}}
\newcommand{\spp}{{\textsc{pp}}}
\newcommand{\sep}{{\textsc{ep}}}
\newcommand{\dg}{\Delta\gamma}
\newcommand{\enf}{\epsilon_{\rm nf}}
\newcommand{\fcme}{f_{\textsc{cme}}}
\newcommand {\mean}[1]   {\langle{#1}\rangle}

\newcommand {\red}[1]   {\textcolor{red}{#1}}
\newcommand {\blue}[1]  {\textcolor{blue}{#1}}
\newcommand {\green}[1] {\textcolor{green}{#1}}


\title{Azimuthal distribution of exponential format for particle collective motions in heavy-ion collisions under asynchronous assumption}

\author{Yicheng Feng}
\email{feng216@purdue.edu}
\address{Department of Physics and Astronomy, Purdue University, West Lafayette, IN 47907, USA}


\date{\today} 


\begin{abstract}
    Particle azimuth distributions are widely studied in heavy-ion collisions. They are often expanded in Fourier series to extract anisotropic flow harmonics simultaneously.
	It was recently proposed that the different orders of flows could happen asynchronously or noninterdependently. 
	This study extends this idea to an exponential format of the azimuth distribution, which makes it straightforward to extract the asynchronous flow coefficients. We compare these new coefficients to the conventional ones, and find consistency in the leading coefficient and discrepancy in higher-order ones.
\end{abstract}



\maketitle


\section{Introduction} \label{sec:introduction}

The azimuth ($\phi$) distribution is widely studied for the produced particles in heavy-ion collisions. It is straightforward to use Fourier series to expand the distribution into different anisotropic flows simultaneously~\cite{Poskanzer:1998yz}
\begin{equation} \label{eq:fourier}
\begin{split}
	& f(\varphi) \equiv \frac{2\pi}{N}\frac{\der N}{\der \phi} 
	= 1 + \sum_{n=1}^{\infty} 2 v_{n} \cos n \varphi + \sum_{n=1}^{\infty} 2 a_{n} \sin n \varphi \,,
\end{split}
\end{equation}
where $N$ is the number of produced particles and $\varphi$ is particle azimuthal angle relative to the reaction plane $\rp$ (\srp) spanned by the impart parameter and beam direction, $\varphi = \phi - \rp$. 
Suppose $\rp$ is known (e.g., in models), the coefficients $v_{n}$ can be measured from the azimuth of produced particles 
\begin{equation} \label{eq:fourier-vn}
	v_{n} = \frac{1}{2\pi} \int_{0}^{2\pi} f(\varphi) \cos n \varphi \der\varphi = \langle \cos n \varphi \rangle \,.
\end{equation}
The coefficients $a_{n}$ may have dependence on the charge of particles to characterize the charge separation phenomena inside one event like the chiral magnetic effect~\cite{Kharzeev:1998kz,Kharzeev:2004ey,Kharzeev:2007jp}. 

Recently, a new study~\cite{Xu:2023psp} proposes that the flows with different orders might happen asynchronously or noninterdependently. 
The $n^{\text{th}}$-order flow operator 
$(1 + 2 v_{n} \cos n \varphi)$ is acted one after another~\cite{Xu:2023psp}
\begin{equation} \label{eq:async}
	f(\varphi) = \prod_{n=1}^{\infty} (1 + 2 \tilde{v}_{n} \cos n \varphi) \prod_{n=1}^{\infty} (1 + 2 \tilde{a}_{n} \sin n \varphi) \,,
\end{equation}
where $\tilde{v}_{n}$ and $\tilde{a}_{n}$ are used to distinguish from the original Fourier coefficients.
Since the operations are commutative, the (time) sequence of different flow terms does not matter. 
In this asynchronous format, the coefficients are hard to extract from particle azimuthal distribution because of cross terms, so only a few orders of $\tilde{v}_{n}$ are discussed at a time with approximations in Ref.~\cite{Xu:2023psp}.



\section{Exponential format} \label{sec:exp}

Besides asynchronous build up of harmonic flows one after the other~\cite{Xu:2023psp}, one may take a step further to purport that each harmonic flow is built up gradually over time. 
Without loss of generality, suppose the flow $\tilde{v}_{n}$ of order $n$ is accumulated over $m$ equal fractions
\begin{equation} \label{eq:mstep}
	\left(1 + \frac{2 \tilde{v}_{n}}{m} \cos n \varphi \right)^{m} \,.
\end{equation}
Letting each fraction go to infinitesimal ($m \rightarrow \infty$), Eq.~\ref{eq:mstep} reduces to exponential
\begin{equation}
	\lim_{m\rightarrow\infty} \left(1 + \frac{2 \tilde{v}_{n}}{m} \cos n \varphi \right)^{m}
	= e^{2 \tilde{v}_{n} \cos n \varphi} \,.
\end{equation}
The equal step size is only to simplify the derivation, but the result is general. 
Considering all coefficients, the asynchronous expansion (Eq.~\ref{eq:async}) can be written into the exponential format
\begin{equation} \label{eq:exp}
	f(\varphi) 
	= A \exp \left\{ \sum_{n=1}^{\infty} 2 \tilde{v}_{n} \cos n \varphi  + \sum_{n=1}^{\infty} 2 \tilde{a}_{n} \sin n \varphi \right\} \,,
\end{equation}
where $A$ is a scaling factor to normalize this distribution $f(\varphi)/(2\pi)$ and its value is usually around 1. 
Note that Eqs.~\ref{eq:async} and \ref{eq:exp} are two different expansions with different coefficients. In the rest of this paper, if not specified, the notations $\tilde{v}_{n}$ and $\tilde{a}_{n}$ stand for the coefficients in Eq.~\ref{eq:exp}.
From this format, it is straightforward to extract the coefficients $\tilde{v}_{n}$ ($n \ge 1$)
\begin{equation} \label{eq:exp-vn}
	\tilde{v}_{n} = \frac{1}{2\pi} \int_{0}^{2\pi} \ln\left[f(\varphi) \right] \cos n \varphi \der \varphi \,,
\end{equation}
where the scaling factor $A$ does not make any contribution.
However, this format cannot be expressed as an average over produced particles. Instead, we need to get the azimuth distribution $f(\varphi)$ first, and then take logarithm and integral.

All those formats are consistent with each other to the leading order, since the coefficients are usually much smaller than $1$. 
We start from Eq.~\ref{eq:exp}, drop the constant factor $A$, and make approximations: 
\begin{equation}
\begin{split}
	f(\varphi) =& \prod_{n=1}^{\infty} e^{2 \tilde{v}_{n} \cos n \varphi} \prod_{n=1}^{\infty} e^{2 \tilde{a}_{n} \sin n \varphi} \\
	\approx& \prod_{n=1}^{\infty} (1 + 2 \tilde{v}_{n} \cos n \varphi) \prod_{n=1}^{\infty} (1 + 2 \tilde{a}_{n} \sin n \varphi) \,,
\end{split}
\end{equation}
where the last line agrees with Eq.~\ref{eq:async} to the first order of each coefficient separately. 
However, the exponential format introduces more higher-order terms.
For example, consider the case with only $\tilde{v}_{n}$, $\tilde{v}_{2n}$, and $\tilde{a}_{n}$
\begin{equation} \label{eq:exp-example}
\begin{split}
	& e^{2 \tilde{v}_{n} \cos n\varphi} e^{2 \tilde{v}_{2n} \cos2n\varphi} e^{2 \tilde{a}_{n} \sin n\varphi} \\
	\approx& \left(1 + 2 \tilde{v}_{n} \cos n\varphi + 2 \tilde{v}_{n}^{2} \cos^{2}n\varphi \right) \\ 
	& \times \left(1 + 2 \tilde{v}_{2n} \cos2n\varphi + 2 \tilde{v}_{2n}^{2} \cos^{2}2n\varphi \right) \\ 
	& \times \left(1 + 2 \tilde{a}_{n} \sin n\varphi + 2 \tilde{a}_{n}^{2} \sin^{2}n\varphi \right) \\
	\approx& (1 + \tilde{v}_{n}^{2} + \tilde{v}_{2n}^{2} + \tilde{a}_{n}^{2}) \\
	& + 2 \tilde{v}_{n} ( 1 + \tilde{v}_{2n}) \cos n \varphi \\ 
	& + (2\tilde{v}_{2n} + \tilde{v}_{n}^{2} - \tilde{a}_{n}^{2}) \cos2n\varphi \\
	& + 2 \tilde{v}_{n} \tilde{v}_{2n} \cos3n\varphi \\
	& + \tilde{v}_{2n}^{2} \cos 4n\varphi \\
	& + 2 \tilde{a}_{n} (1 - \tilde{v}_{2n}) \sin n \varphi \\
	& + 2 \tilde{a}_{n} \tilde{v}_{n} \sin2n\varphi \\
	& + 2 \tilde{a}_{n} \tilde{v}_{2n} \sin3n\varphi \,,
\end{split}
\end{equation}
where the squared terms do not exist in the similar expansion of Eq.~\ref{eq:async}. 
If we set $n=1$, then we can see the term $\tilde{v}_{1} \tilde{v}_{2}$ contributes to $v_{3}$ and $\tilde{a}_{1} \tilde{v}_{2}$ contributes to $a_{3}$, as also shown by Ref.~\cite{Xu:2023psp}. 

Another approximation of Eq.~\ref{eq:exp} is 
\begin{equation}
\begin{split}
	f(\varphi) 
	=& \exp \left\{ \sum_{n=1}^{\infty} 2 \tilde{v}_{n} \cos n \varphi + \sum_{n=1}^{\infty} 2 \tilde{a}_{n} \sin n \varphi \right\} \\
	\approx& 1 + \sum_{n=1}^{\infty} 2 \tilde{v}_{n} \cos n \varphi + \sum_{n=1}^{\infty} 2 \tilde{a}_{n} \sin n \varphi \,,
\end{split}
\end{equation}
so roughly $v_{n} \approx \tilde{v}_{n}$ and $a_{n} \approx \tilde{a}_{n}$ to the leading order by comparison to Eq.~\ref{eq:fourier}.

\begin{figure}[h]
	\includegraphics[width=0.8\linewidth]{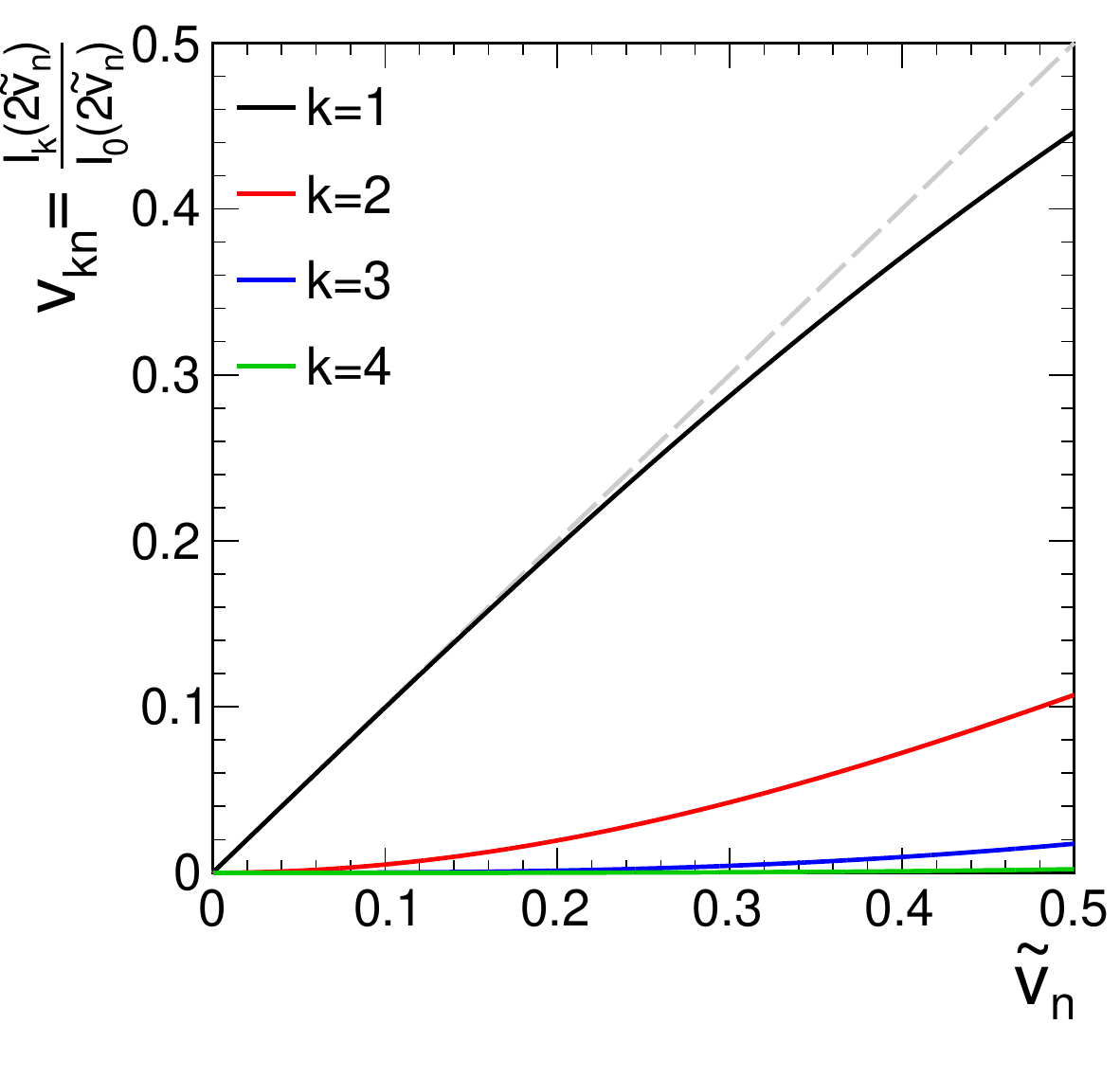}
	\caption{The modified Bessel functions for the relationships between $v_{nk}$ and $\tilde{v}_{n}$ (Eq.~\ref{eq:bessel}) with the simplified exponential format of azimuth distribution (Eq.~\ref{eq:exp:fn}). For $k=1$, the black curve follows the dashed diagonal ($v_{n} = \tilde{v}_{n}$) at small $\tilde{v}_{n}$ (typically a few percent), while all curves with higher $k$ are relatively small.}
	\label{fig:bessel}
\end{figure}

As a check, take a simplified case with only order $n$
\begin{equation} \label{eq:exp:fn}
	f_{n}(\varphi) \equiv A_{n} e^{2 \tilde{v}_{n} \cos n \varphi} \,,
\end{equation}
then from Eq.~\ref{eq:fourier-vn}
\begin{equation} \label{eq:bessel}
\begin{split}
	v_{k n} =& \int_{0}^{2\pi} f_{n}(\varphi) \cos k n \varphi \der \varphi \left/ \int_{0}^{2\pi} f_{n}(\varphi) \der\varphi \right. \\
	=& \int_{0}^{2\pi} e^{2\tilde{v}_{n} \cos n\varphi} \cos k n \varphi \der \varphi \left/ \int_{0}^{2\pi} e^{2\tilde{v}_{n} \cos n\varphi} \der\varphi \right. \\
	=& I_{k}(2 \tilde{v}_{n}) \left/ I_{0}(2\tilde{v}_{n}) \right. \,,
\end{split}
\end{equation}
where $I_{0}(x)$ and $I_{k}(x)$ are the modified Bessel function of the first kind with integer order $0$ and $k \ge 1$ respectively~\cite{Wiki:BesselFunction}. 
The asymptotic form of modified Bessel functions at small $x$ is $I_{k}(x) \sim \frac{1}{k!} \left(\frac{x}{2}\right)^{k}$, so
\begin{equation} \label{eq:asymptotic}
	v_{kn} 
	\sim \frac{1}{k!} \tilde{v}_{n}^{k} \,.
\end{equation}
It is clear that $v_{n} \sim \tilde{v}_{n}$ only for $k=1$ and all other $k$ gives small $v_{kn}$ of higher order, which are shown in Fig.~\ref{fig:bessel}. 
Note that the discussions here only cover some of the Fourier coefficients $v_{kn}$ of the simplified exponential format $f_{n}(\varphi)$. For all Fourier coefficients of the general exponential format (Eq.~\ref{eq:exp}), the analytical calculations are difficult and out of the scope of this paper. Instead, we will use model study and numerical calculation in the next section.

This exponential expansion (Eq.~\ref{eq:exp}) is convenient in studies of the number-of-constituent-quark (NCQ) scaling~\cite{Molnar:2003ff,STAR:2003wqp,PHENIX:2003qra}. 
For a hadron made of $n_{q}$ quarks, its azimuthal distribution can be expressed in the simple quark coalescence model~\cite{Molnar:2003ff,Sato:1981ez,Schwarzschild:1963zz} by 
\begin{equation}
\begin{split}
	f_{h}(\varphi; p_{T}) \propto & \left[ f_{q}\left(\varphi; \frac{p_{T}}{n_{q}}\right) \right]^{n_{q}} \,.
\end{split}
\end{equation}
The conventional expansion (Eq.~\ref{eq:fourier}) for NCQ scaling is
\begin{equation}
\begin{split}
	&1 + \sum_{n=1}^{\infty} 2 v_{n,h}\left(p_{T}\right) \cos n \varphi \\
	&\propto \left[ 1 + \sum_{n=1}^{\infty} 2 v_{n,q}\left(\frac{p_{T}}{n_{q}}\right) \cos n \varphi \right]^{n_{q}} \,,
\end{split}
\end{equation}
while the exponential expansion (Eq.~\ref{eq:exp}) gives
\begin{equation}
\begin{split}
	& \exp\left\{ \sum_{n=1}^{\infty} 2 \tilde{v}_{n,h}\left(p_{T}\right) \cos n \varphi \right\} \\ 
	& \propto \left[\exp\left\{ \sum_{n=1}^{\infty} 2 \tilde{v}_{n,q}\left(\frac{p_{T}}{n_{q}}\right) \cos n \varphi \right\} \right]^{n_{q}} \,,
\end{split}
\end{equation}
where $a_{n}$ ($\tilde{a}_{n}$) terms are dropped and $v_{n}$ ($\tilde{v}_{n}$) are regarded as functions of transverse momentum $p_{T}$. The subscripts $h$ and $q$ stand for hadron and quark respectively. 
The exponential format conveniently brings the constituent quark number $n_{q}$ in front of $\tilde{v}_{n}$, $\tilde{v}_{n,h}(p_{T}) = n_{q} \tilde{v}_{n,q}(p_{T}/n_{q})$, whereas the Fourier format needs approximation, $v_{n,h}(p_{T}) \approx n_{q} v_{n,q}(p_{T}/n_{q})$ with higher-order terms dropped.


\section{AMPT model simulation} \label{sec:ampt}

A Multiphase Transport (\ampt) model~\cite{Zhang:1999bd,Lin:2004en} version v1.25t4cu2/v2.25t4cu2 is used to simulate Au+Au collisions at $\snn = 200$ GeV, in which string melting is implemented and charge conservation is ensured.
In those simulations, hadronic cascade is included by setting the parameter $\tt{NTMAX=150}$. 
This version of \ampt\ can describe anisotropic flow and particle production data~\cite{Lin:2001zk, Lin:2014tya}, and reasonably describe the measured correlations~\cite{Bzdak:2014dia}.
In this study, about 22 million minimum-bias events are simulated with impart parameter $0 \le b \le 16$ fm. The particles of interest (\poi) are the final-state protons, pions, and kaons with transverse momentum $0.2 < p_{T} < 2 \text{ GeV}/c$ and pseudorapidity $-1<\eta<1$. According to the \poi\ multiplicity distribution, the centrality is defined and the range 0--80\% is used.

\begin{figure}
	\includegraphics[width=1.0\linewidth]{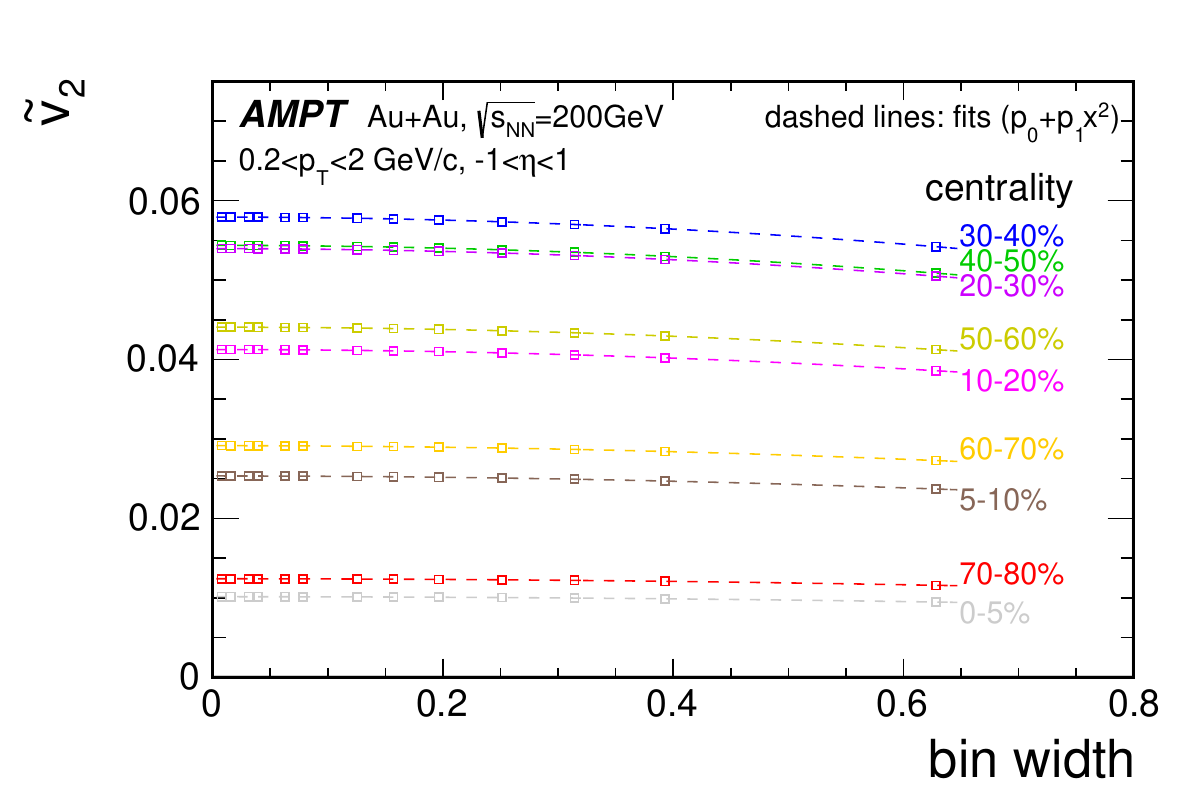}
	\caption{The $\tilde{v}_{2}$ coefficients as functions of bin width for various centrality ranges. The statistical uncertainties are too small to be visible. The dashed lines are fits ($p_{0}+p_{1} x^{2}$) for extrapolation.}
	\label{fig:exp-v2-bw}
\end{figure}

\begin{figure}
	\includegraphics[width=1.0\linewidth]{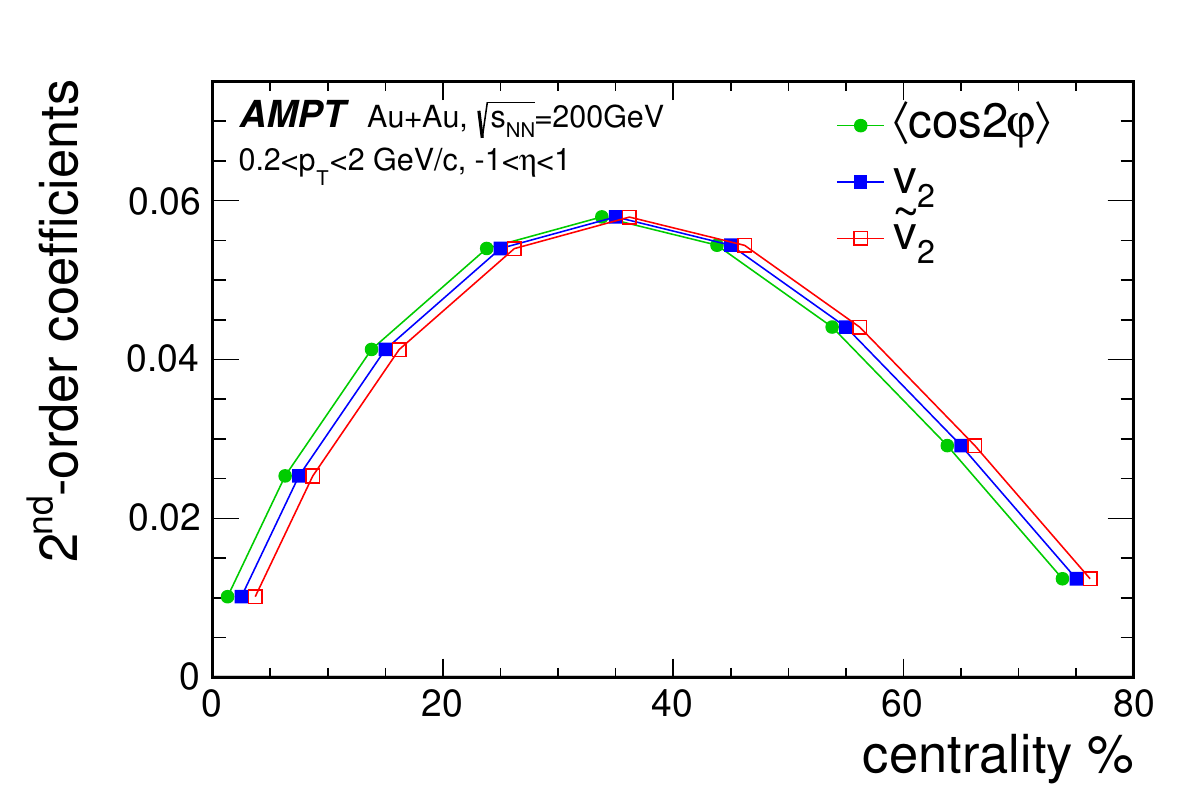}
	\caption{The second-order coefficients as functions of centrality. The closed and open squares are the extrapolated coefficients $v_{2}$ and $\tilde{v}_{2}$ from $f(\varphi)$ histograms, while the closed circle is the track-level average $\langle \cos2\varphi \rangle$. The data points are shifted along $x$-axis for clarity.}
	\label{fig:v2}
\end{figure}

We will try to calculate the flow coefficients of the second order ($v_{2}$, $\tilde{v}_{2}$) and fourth order ($v_{4}$, $\tilde{v}_{4}$) for the original Fourier expansion and the exponential format.
The \srp\ azimuth is fixed to $\rp = 0$ in \ampt\ model. 
As mentioned after Eq.~\ref{eq:exp-vn}, $\tilde{v}_{n}$ can only be calculated from the $\varphi$ distribution $f(\varphi)$, but in practice only histograms with finite bin width are available. The finite bin size underestimate the coefficients, effectively like the resolution effect. 
To circumvent this, we calculate the $\tilde{v}_{n}$ value from $f(\varphi)$ histograms with different bin width, and extrapolate the value to vanishing bin width. Figure~\ref{fig:exp-v2-bw} shows such extrapolations for $\tilde{v}_{2}$, where it is clear that the effect is already small when the bin width is less than $0.02\pi$ (e.g., 100 bins over $2\pi$). 

\begin{figure}
	\includegraphics[width=1.0\linewidth]{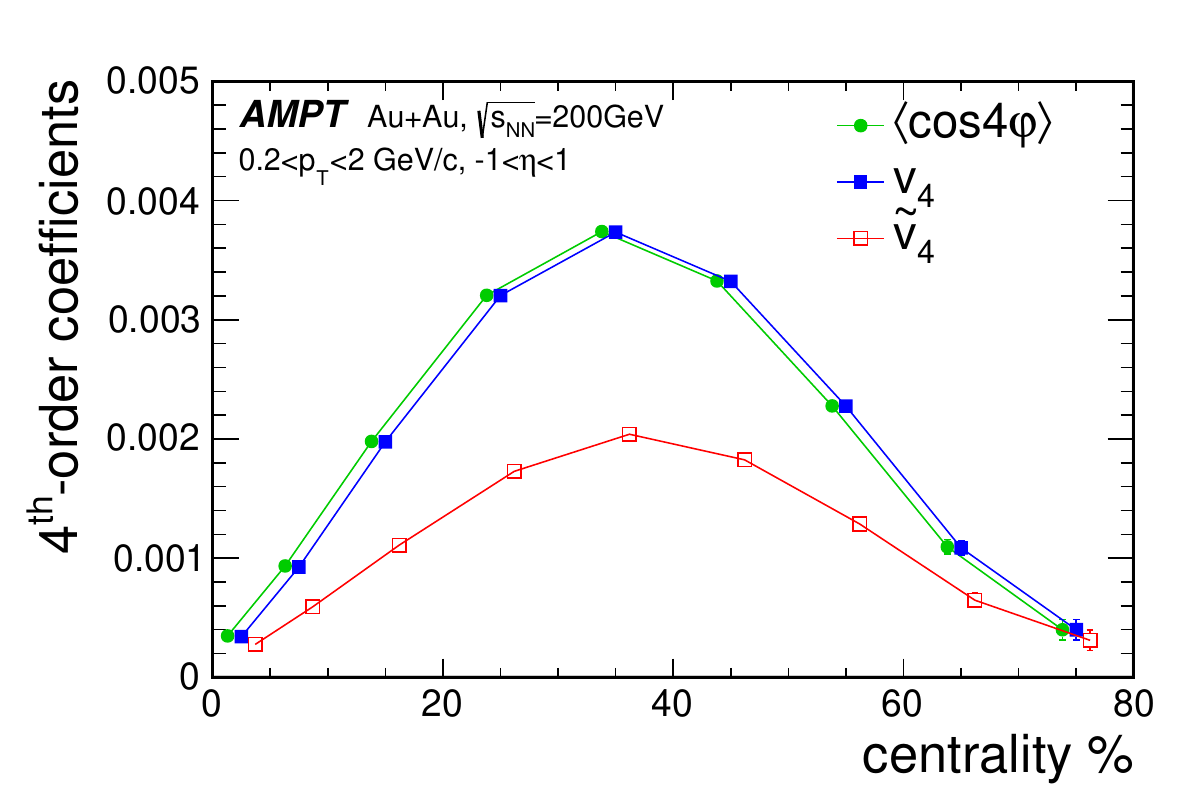}
	\caption{The fourth-order coefficients as functions of centrality. The closed and open squares are the extrapolated coefficients $v_{4}$ and $\tilde{v}_{4}$ from $f(\varphi)$ histograms, while the closed circle is the track-level average $\langle \cos4\varphi \rangle$. The data points are shifted along $x$-axis for clarity.}
	\label{fig:v4}
\end{figure}

\begin{figure}
	\includegraphics[width=1.0\linewidth]{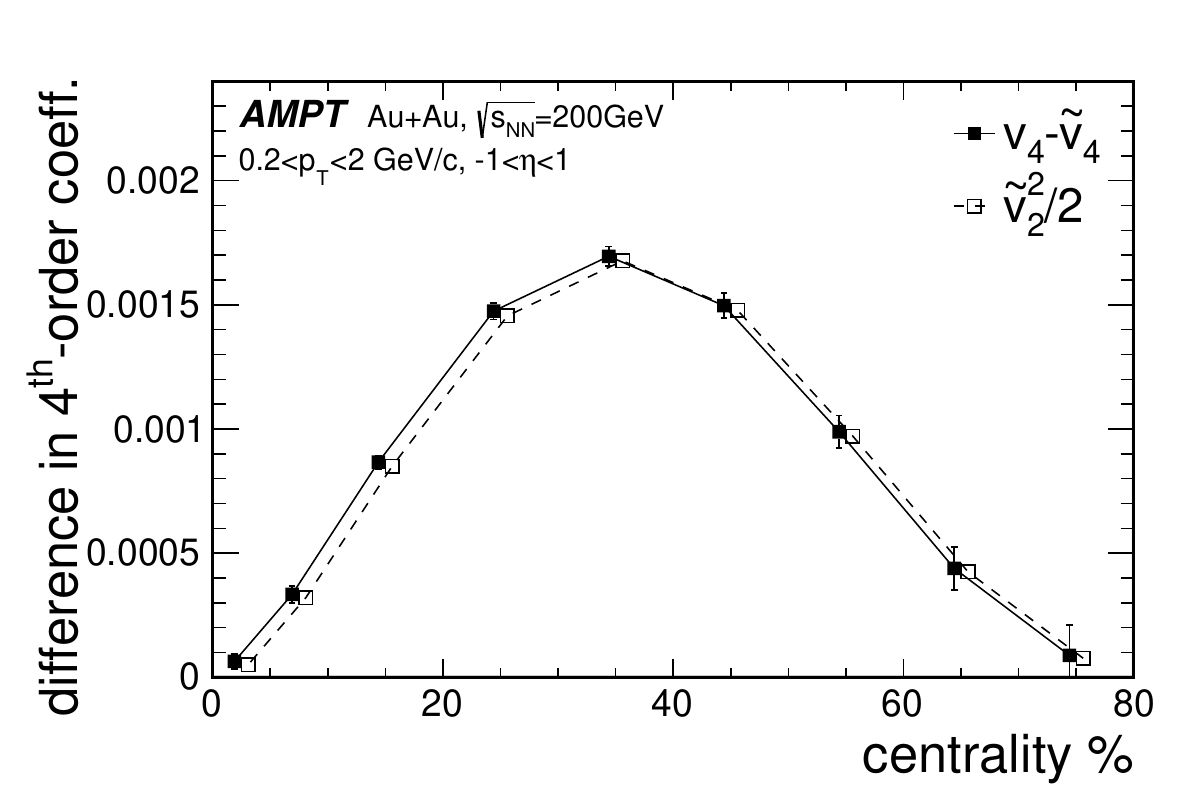}
	\caption{The differences between $v_{4}$ and $\tilde{v}_{4}$ as functions of centrality (solid line). The dashed line $\tilde{v}_{2}^{2}/2$ is the leading-order contribution to $v_{4}$ from $\tilde{v}_{2}$ (Eq.~\ref{eq:asymptotic}). }
	\label{fig:difv4}
\end{figure}

Figure~\ref{fig:v2} shows the second-order coefficients as functions of centrality from different methods. The three curves, extrapolated $v_{2}$, $\tilde{v}_{2}$, and track average $\langle \cos2\varphi \rangle$, are nearly identical to each other--the relative difference is less than $0.02\%$. 
Figure~\ref{fig:v4} shows the fourth-order coefficients, where $v_{4}$ and $\tilde{v}_{4}$ are obviously different. This difference comes from the contribution from $\tilde{v}_{2}$ to $v_{4}$, $I_{2}(2\tilde{v}_{2})/I_{0}(2\tilde{v}_{2}) \approx \tilde{v}_{2}^{2}/2$ (Eqs.~\ref{eq:bessel}, \ref{eq:asymptotic}), which can also be seen from the example expansion in Eq.~\ref{eq:exp-example} (coefficient for $\cos2n\varphi$ term with $n=2$ and $\tilde{a}_{n}=0$). Figure~\ref{fig:difv4} shows $\tilde{v}_{2}^{2}/2$ can describe most of $v_{4}-\tilde{v}_{4}$, and the residue may come from higher-order terms of $\tilde{v}_{2}$ or contributions from other coefficients.
Since the magnitude of $v_{4}$ is usually much smaller than $v_{2}$ and around $v_{2}^{2}$, the $\tilde{v}_{2}$ contribution is significant. 

This begs the question which one is more basic, the $v_{n}$ or $\tilde{v}_{n}$. This question has to be answered by physics, as obviously any azimuthal distribution can mathematically be described by multiple expansions (Eqs.~\ref{eq:fourier}, \ref{eq:async}, \ref{eq:exp}). 
For instance, the charge separation from the chiral magnetic effect~\cite{Kharzeev:1998kz,Kharzeev:2004ey,Kharzeev:2007jp} happens at early stage before the development of flows, so it may be more reasonable to consider $a_{1}$ and $v_{n}$ as asynchronous terms. 
Flow responds to the shape of the collision zone, so it is intuitive to use simultaneous Fourier expansion for the ${v}_{n}$ terms. 
On the other hand, each flow component is developped over the time evolution of the collision system, so the exponential expression of Eq.~\ref{eq:exp} seems more appropriate. 
It is interesting to note that the exponential format bridges the Fourier and asynchronous formats--different terms act on each other by products, while the same Fourier format appears as the exponent.
These physics investigations of the origins of the harmonic flow coefficients are beyond the scope of the present work, and hopefully will simulate future studies.


\section{Summary} \label{sec:summary}

This study is motivated by the work of asynchronous flow~\cite{Xu:2023psp} and takes a step further to consider the development of flow of each harmonic order over time evolution of the collision system.
We arrive at an exponential format of the expansion series for the azimuth distribution of particles. 
The new coefficients $\tilde{v}_{n}$ can be readily calculated with the exponential format. 
They can be used in the NCQ scaling. 
 
The relationship between the original Fourier coefficients $v_{n}$ and the new ones $\tilde{v}_{n}$ from this exponential format are studied by the analytical calculations and \ampt\ model simulations. It is found that $v_{kn} = I_{k}(2\tilde{v}_{n}) / I_{0}(2\tilde{v}_{n}) \approx \tilde{v}_{n}^{k} / k!$ with integer $k \ge 1$ for a simplified distribution $\propto e^{2 \tilde{v}_{n} \cos n \varphi}$.
This suggests $\tilde{v}_{n}$ mainly contributes to $v_{n} \sim \tilde{v}_{n}$, while its contribution to other terms like $v_{kn}$ ($k > 1$) is of higher order, $\tilde{v}_{n}^{k}$.
\ampt\ results show that $v_{2}$ agrees with $\tilde{v}_{2}$ well, but $v_{4}$ is larger than $\tilde{v}_{4}$. This is because $v_{4}$ itself is already small, so the second-order contribution, $\tilde{v}_{2}^{2}/2$, can still make significant difference.



\section*{Acknowledgment} 

The author thanks Gang Wang, Zhiwan Xu, Aihong Tang, and Fuqiang Wang for fruitful discussions. 
This work is supported by the U.S.~Department of Energy (Grant No.~DE-SC0012910).


\bibliography{./ref}


\end{document}